%% file: draft.tex
\documentclass[sigconf]{acmart}
\AtBeginDocument{%
  \providecommand\BibTeX{{%
    \normalfont B\kern-0.5em{\scshape i\kern-0.25em b}\kern-0.8em\TeX}}}

\setcopyright{acmcopyright}
\copyrightyear{2024}
\acmYear{2024}
\setcopyright{rightsretained}
\acmConference[CHI EA '24]{Extended Abstracts of the CHI Conference on Human Factors in Computing Systems}{May 11--16, 2024}{Honolulu, HI, USA}
\acmBooktitle{Extended Abstracts of the CHI Conference on Human Factors in Computing Systems (CHI EA '24), May 11--16, 2024, Honolulu, HI, USA}
\acmDOI{10.1145/3613905.3650786}
\acmISBN{979-8-4007-0331-7/24/05}

\usepackage{multirow}
\usepackage{colortbl}
\usepackage{xspace}
\usepackage{graphicx}
\usepackage{xcolor}
\usepackage{subcaption}
\usepackage{soul}

\newif\ifcomment

\ifcomment
  \newcommand{\gao}[1]{\textcolor[rgb]{0.6,0,0.2}{Jie: #1}}
  \newcommand{\simret}[1]{\textcolor[rgb]{0.5,0.5,0.2}{Simret: #1}}
  \newcommand{\added}[1]{\textcolor[rgb]{0.2,0.5,0.8}{#1}}
  \newcommand{\deleted}[1]{\textcolor[rgb]{0.8,0.8,0.8}{#1}}
\else
  \newcommand{\gao}[1]{}
  \newcommand{\simret}[1]{}
  \newcommand{\added}[1]{\textcolor{black}{#1}}
  \newcommand{\deleted}[1]{}
\fi

\newcommand{\standard}{{\textit{Mode 1: Standard Prompting}}\xspace}
\newcommand{\uiinterface}{{\textit{Mode 2: User Interface}}\xspace}
\newcommand{\context}{{\textit{Mode 3: Context-based}}\xspace}
\newcommand{\agent}{{\textit{Mode 4: Agent Facilitator}}\xspace}

\begin{document}

\title{A Taxonomy for Human-LLM Interaction Modes: An Initial Exploration}

\author{Jie Gao}
\authornote{Both authors contributed equally to this paper.}
\email{jie.gao@smart.mit.edu}
\affiliation{%
    \institution{Singapore-MIT Alliance for Research and Technology}
    \country{Singapore}
    }

\author{Simret Araya Gebreegziabher}
\authornotemark[1]
\email{sgebreeg@nd.edu}
\affiliation{%
    \institution{University of Notre Dame}
    \city{Notre Dame}
    \state{Indiana}
    \country{USA}
    }

\author{Kenny Tsu Wei Choo}
\email{kenny@kennychoo.net}
\affiliation{
    \institution{Singapore University of Technology and Design}
   \country{Singapore}}

\author{Toby Jia-Jun Li}
\email{toby.j.li@nd.edu}
\affiliation{
    \institution{University of Notre Dame}
    \city{Notre Dame}
    \state{Indiana}
    \country{USA}
    }

\author{Simon Tangi Perrault}
\email{perrault.simon@gmail.com}
\affiliation{
    \institution{Singapore University of Technology and Design}
   \country{Singapore}}

\author{Thomas W. Malone}
\email{malone@mit.edu}
\affiliation{
    \institution{Massachusetts Institute of Technology}
    \city{Cambridge}
    \state{Massachusetts}
   \country{USA}}

\renewcommand{\shortauthors}{Jie Gao, Simret Araya Gebreegziabher et al.}


\input{paper/0_abstract}







\maketitle

\input{paper/1_introduction}
\input{paper/3_method}
\input{paper/4_link_taxonomy}

\input{paper/5_limitation}
\input{paper/6_conclusion}

% \begin{acks}
% \end{acks}

\bibliographystyle{ACM-Reference-Format}
\bibliography{draft}
% \bibliography{paper/CSCW-IUI-UIST}

\newpage
\input{paper/appendix}

\end{document}

%% file: paper/0_abstract.tex
\begin{abstract}
    With ChatGPT's release, conversational prompting has become the most popular form of human-LLM interaction. However, its effectiveness is limited for more complex tasks involving reasoning, creativity, and iteration. Through a systematic analysis of HCI papers published since 2021, we identified four key phases in the human-LLM interaction flow—\textit{planning}, \textit{facilitating}, \textit{iterating}, and \textit{testing}—to precisely understand the dynamics of this process. Additionally, we have developed a taxonomy of four primary interaction modes: \standard, \uiinterface, \context, and \agent. This taxonomy was further enriched using the ``5W1H'' guideline method, which involved a detailed examination of definitions, participant roles (Who), the phases that happened (When), human objectives and LLM abilities (What), and the mechanics of each interaction mode (How). We anticipate this taxonomy will contribute to the future design and evaluation of human-LLM interaction.
\end{abstract}

%% file: paper/1_introduction.tex
\section{Introduction and Background}

\begin{table*}[!t]
\centering
\caption{Data Collection. The table shows the initial count of papers at Stage 1 and the number of papers remaining at Stage 2 after the filtering process.}
\scalebox{0.85}{\begin{tabular}{@{}l|ccc|ccc|ccc|ccc|c@{}}
\toprule
& \rotatebox{0}{CHI'23} & \rotatebox{0}{CHI'22} & \rotatebox{0}{CHI'21} & \rotatebox{0}{UIST'23} & \rotatebox{0}{UIST'22} & \rotatebox{0}{UIST'21} & \rotatebox{0}{CSCW'23} & \rotatebox{0}{CSCW'22} & \rotatebox{0}{CSCW'21} & \rotatebox{0}{IUI'23} & \rotatebox{0}{IUI'22} & \rotatebox{0}{IUI'21} & Total \\ 
\midrule
\rowcolor[HTML]{EFEFEF} 
\begin{tabular}[c]{@{}l@{}}Stage 1: Initial Searching \end{tabular} & 43 & 23 & 19 & 12 & 8 & 7 & 11 & 5 & 10 & 9 & 5 & 12 & \textit{164} \\
\begin{tabular}[c]{@{}l@{}}Stage 2: (Post) Paper Filtering \end{tabular} & 19 & 10 & 5 & 11 & 4 & 3 & 4 & 2 & 4 & 6 & 3 & 2 & \textit{73} \\
\bottomrule
\end{tabular}}
\label{tab:paper_count}
\end{table*}

\added{Every use of computers involves an \textit{interaction mode}---a pattern of interaction between the user and the computer. This concept has evolved significantly, starting from command-line interfaces on early teletypes, advancing to the direct manipulation of on-screen images, and progressing to engaging conversations with chatbots, among others~\cite{designing2004shneiderman}.}

Since the introduction of Large Language Models (LLMs), especially ChatGPT, conversational interactions have become the "default" interaction mode for the interaction between human users and LLMs. \added{This extends to other notable platforms like Claude\footnote{\url{https://claude.ai/chats}}, Gemini\footnote{\url{https://gemini.google.com/app}}, and Llama 2\footnote{\url{https://www.llama2.ai/}}, among others.}
\added{This interaction is based on prompts—specific instructions given to an LLM that allow it to grasp the user's intent and then generate meaningful outcomes through dialogue. Thus, the creation of high-quality, well-considered prompts for the dialogue is critical for enhancing outcomes.} 

\added{Moreover, intricately designed prompts enable the execution of complex tasks, such as solving mathematical problems, writing, and coding. The design strategies include zero-shot~\cite{kojima2022large}, few-shot~\cite{logan2021cutting}, chain-of-thought techniques~\cite{wei2022chain}, etc.}
\added{Furthermore, White et al.~\cite{white2023prompt} have identified a series of prompt patterns—similar to software design patterns~\cite{gamma1995design, schmidt2013pattern}—that can be employed to construct complex prompts.} For example, the "flipped interaction" pattern, where LLMs initiate questions instead of merely producing outputs; the "gameplay pattern", generating output in game format; and the "infinite generation pattern", enabling continuous output generation without repeated user prompts.

\added{Recently, HCI researchers have pushed the boundaries of conversational interaction capabilities through diverse human-LLM interaction designs. The key strategy is to leverage various HCI techniques like visual programming and direct manipulation, to develop complex prompts. These prompts significantly enhance LLMs, equipping them with sophisticated capabilities for complex tasks, including argumentative writing~\cite{zhang2023visar}, brainstorming~\cite{jiang2023graphologue, suh2023sensecape}, and more.} However, they primarily focus on developing specific interaction modes, often overlooking a more comprehensive framework that encompasses various interaction perspectives. Exploring the various types of interaction modes within such a framework is promising, yet remains largely underexplored.

In this work, we investigate the following research questions about the human-LLM interaction:

\begin{figure*}[!t]
    \centering
    \includegraphics[width=0.65\linewidth]{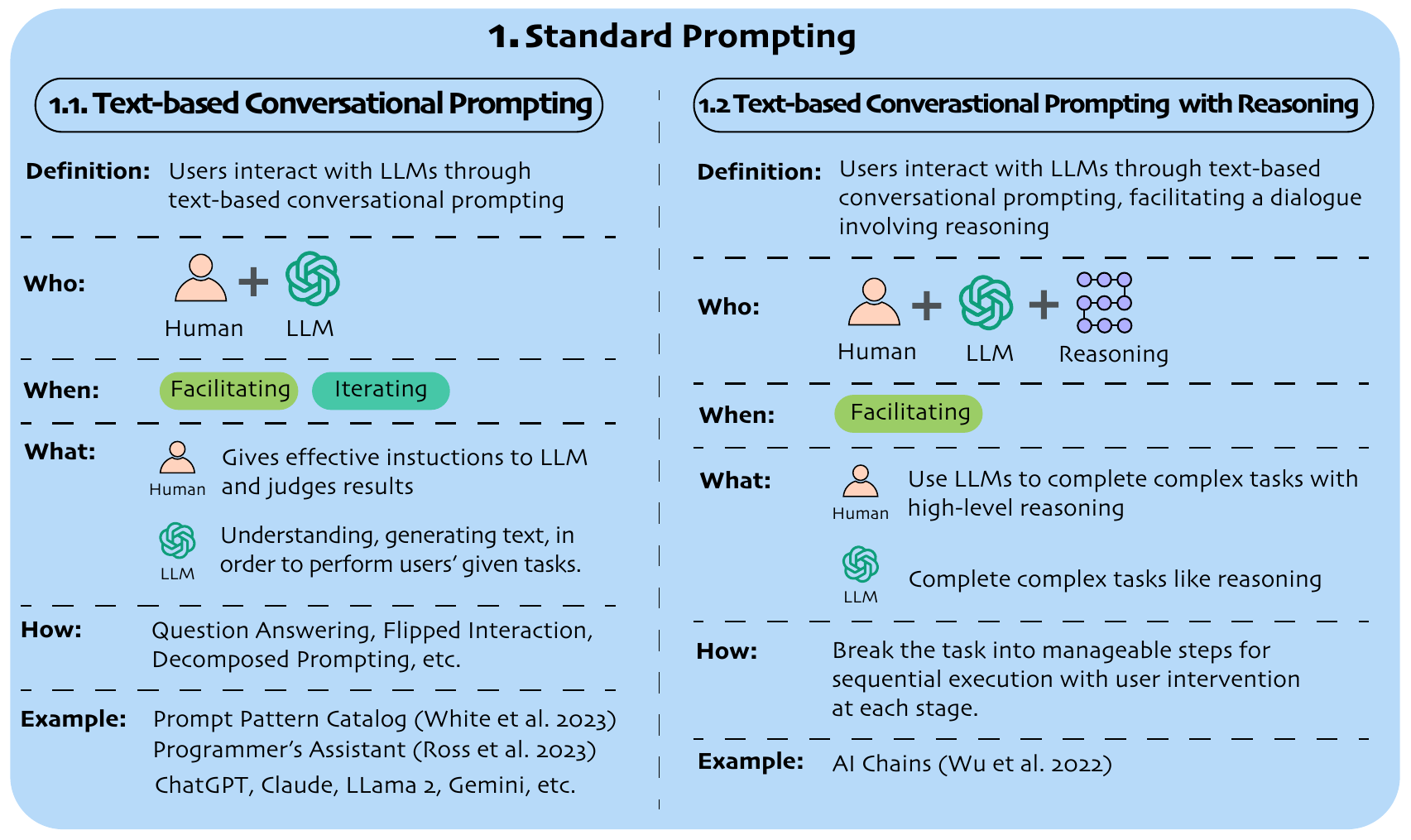}
    \caption{\standard. }
    \label{fig:taxonomy1}
\end{figure*}

\begin{itemize}
\item \textit{RQ1: What are the different phases in a human-LLM interaction flow?} Although much research has started to reference "human-LLM interaction" \footnote{See \textbf{\textit{From Thought to Prompt: Cognitive Design Challenges in Human-LLM Interactions:}} \url{https://www.aalto.fi/en/events/from-thought-to-prompt-cognitive-design-challenges-in-human-llm-interactions}, \textbf{\textit{Call for papers for human-LLM interaction:}} \url{https://human-llm-interaction.github.io/workshop/hri24/call-for-papers}, \textbf{\textit{Low-code LLM}}~\cite{cai2023lowcode}, and many others.}, a precise definition of this term remains vague. What exactly does it encompass? Therefore, gaining a more accurate understanding of its nature, including the phases within a ``Human-LLM Interaction" flow, is crucial.
\item \textit{RQ2: How can the various interaction modes between humans and LLMs be categorized? Is it possible to structure these into a taxonomy?}
\end{itemize}

To answer our research questions, we performed a systematic review of existing literature in HCI venues published since 2021, including CHI, CSCW, UIST, and IUI. Our analysis of current literature on LLM-powered tools and systems has led to the identification of four crucial temporal phases in these interaction flows, namely \textit{planning}, \textit{facilitating}, \textit{iterating}, and \textit{testing}. Additionally, our research introduces a detailed, structured taxonomy that encapsulates four modes of interaction between humans and LLMs, including \standard, \uiinterface, \context, and \agent. \added{We anticipate that these interaction modes, initially foundational in writing and coding, will become crucial across various tasks where prompts act as the primary mechanism driving system functionality, e.g., in image and video generation. As this paper begins an exploration, we anticipate its scope will broaden with the evolving applications of LLMs, thereby influencing the future of human-LLM interaction.}

%% file: paper/3_method.tex
\section{Method}

\subsection{Data Collection}

Our data collection methodology is informed by a systematic literature review protocol~\cite{keele2007guidelines}, which focuses on manual searches across main HCI venues, including conference proceedings like CHI, CSCW, UIST, and IUI. The collection is conducted through two stages.

\subsubsection{Stage 1: Initial Searching} Two authors focused on papers that mentioned key terms including \textit{``Large Language Models''}, \textit{``LLMs''}, \textit{``Natural Language''}, \textit{``prompt/prompting''}, \textit{``generative AI''}, \textit{``GPT''} and \textit{``human-AI''}. The inclusion criteria primarily targeted papers that 1) proposed new platforms or software integrating LLMs, 2) introduced novel interaction techniques with LLMs; and 3) included a few studies on AI agents, which, although not using LLMs, are still relevant to the topic. Moreover, our focus was on research published after 2021, aligning with the period when LLMs gained widespread popularity. In total, there were 164 papers were identified. Table \ref{tab:paper_count} presents the number of papers under each venue.

\subsubsection{Stage 2: Paper Filtering} After collecting the papers, we have performed a deep filtering process, aiming to focus on the most relevant contributions.
\added{The two authors independently summarized and evaluated the collected papers, noting their relevance and contributions, and documented the rationale behind their pertinence.} Following this, the authors convened to discuss their findings and collectively decide on the inclusion or exclusion of each paper. In total, 73 papers were left.

\subsection{Constructing the Taxonomy}

\subsubsection{Stage 1: Development of A Primary Taxonomy}
\label{sec:primary_taxonomy}
Initially, we focused on the types of interactions each paper described. After reviewing and discussing the literature, primary categories were identified (see Appendix Table \ref{tab:primary_taxonomy}). For each category, we adapted the “5W1H” guideline~\footnote{\url{https://ipma.world/5ws-1h-a-technique-to-improve-project-management-efficiencies/}}, and ultimately, we found that only four out of the five dimensions are necessary:

\begin{itemize}
    \item \textbf{Who}: the participated roles in the human-LLM interaction, aiming to understand who is involved and the tasks they perform within the interaction flow.
    \item \textbf{What}: the objectives of \textit{human} engagement and the advanced capabilities \textit{LLMs} gain through augmentation.
    \item \textbf{When}: the phases at which LLM capabilities manifest during the interaction flow.
    \item \textbf{How}: the underlying mechanisms and methods of these interactions.
\end{itemize}

\subsubsection{Stage 2: Systematic Annotation of Each Paper and Refinement of the Primary Taxonomy}

\added{After initially developing the basic taxonomy, the two authors independently annotated more papers, incorporating more categories and detailed dimensions into the taxonomy. Next, they had iterative discussions to address ambiguities and discrepancies, gradually refining the classification criteria for greater clarity.} Lastly, with the refined and more structured taxonomy in hand, the authors proceeded to code the remaining papers for the complete classification (see Appendix Table~\ref{tab:paper_classification}). Through this iterative process, we have developed a taxonomy that captures the intricate details and various types of current human-LLM interactions.

%% file: paper/4_link_taxonomy.tex
\section{Results}

\subsection{RQ1: Four phases of a human-LLM interaction flow}
\added{During the development of the taxonomy, we gained a clearer understanding of the four phases in which LLM assistance typically occurs, including \textit{planning}, \textit{facilitating}, \textit{iterating}, and \textit{testing}.}

\begin{enumerate}
    \item \textit{Planning (before an interaction)}: This phase includes strategizing the entire interaction beyond basic conversational exchanges. In this stage there is a focus to determine the goals of the interaction, and the steps needed to achieve these goals (e.g., determining specific inputs and outputs for each step).
    \item \textit{Facilitating (during an interaction)}: \added{This phase, perhaps the most prevalent, involves assisting users in formulating or completing their interaction proposals, such as text completion. An additional example includes users refining their prompts by asking more in-depth questions or incorporating extra details during conversations with LLMs to achieve their objectives. Furthermore, users evaluate the LLM's varied suggestions, ultimately accepting or rejecting these proposals to fulfill their goals.}
    \item \textit{Iterating (refining an established interaction)}: \added{After establishing and completing an initial interaction flow, users refine and enhance this process through successive adjustments, eliminating the need for further conversation.} This can involve iterating over the \added{existing} prompts and instructions or the outputs through different affordances.
    \item \textit{Testing (testing a defined interaction)}: \added{This phase generates and evaluates diverse responses to variations of user-designed prompts. Such testing is crucial for understanding the breadth and depth of the interactions.}
\end{enumerate}

\subsection{RQ2: Taxonomy}

\begin{figure*}[!t]
    \centering
    \includegraphics[width=0.9\linewidth]{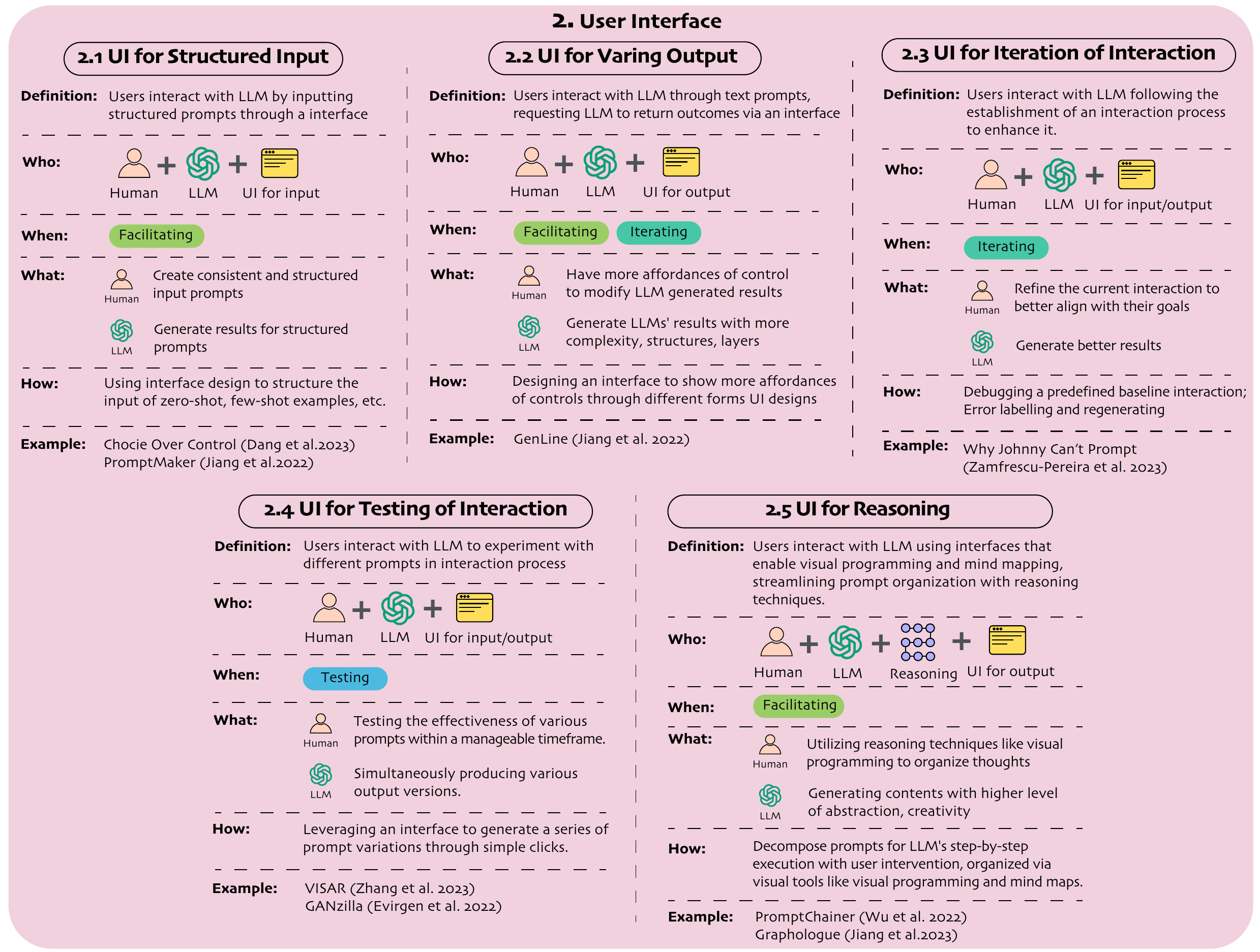}
    \caption{\uiinterface.}
    \label{fig:taxonomy2}
\end{figure*}

\subsubsection{\textbf{Mode 1: Standard Prompting}} Standard Prompting (Figure \ref{fig:taxonomy1}) represents the foundational \added{and widely adopted} interaction mode between humans and LLMs. 

\paragraph{\textbf{Mode 1.1. Text-based Conversational Prompting}}
This mode employs a standard, conversational approach where users design and input prompts, to which the LLM responds with textual outputs~\cite{zhu_leveraging_2023}. This include single-turn conversation~\cite{srinivasa_ragavan_gridbook_2022} and multiple-turn conversations~\cite{bursztyn_developing_2021}. \added{Through each interaction users dynamically elicit responses from the LLM to steer the conversation towards achieving their desired outcomes.} Therefore, both the user's prompts and the LLM's responses evolve to more precisely address the task at hand (such as seeking answers or ideation).

In fact, many platforms, including ChatGPT, Claude, LLama 2, and Gemini, employ this conversational prompting method. From the literature, Programmer's Assistant~\cite{ross_programmers_2023} provides an interface that allows users to interact with the model conversationally. This facilitates understanding code and generating alternative responses through conversational interactions during coding. Another method employs conversational prompting, which can be executed in a "single-turn conversation", involving inputting text into an LLM, which then returns suggestions or completes the text. One example is the OpenAI Playground; similarly, literature such as CoAuthor leverages LLMs to offer writing suggestions based on the user's input text~\cite{lee_coauthor_2022}.

\added{However, this mode often has limitations, as it allows users to input only a limited amount of information through single or multiple-turn prompts. Executing higher-level tasks, such as planning and testing multiple variations, can be challenging. Furthermore, it may also be susceptible to ambiguity and misalignment in interpreting intent~\cite{aina2021language}. Therefore, in the original conversational prompting, prompts must be strategically designed in various ways to further discern user intent. This includes flipped interactions~\cite{aina2021language} or gameplay interaction, etc.~\cite{white2023prompt}.}

\paragraph{\textbf{Mode 1.2. Text-based Conversational Prompting with Reasoning}}
\label{sec:human+llm+reasoning}

\added{While linear conversational interaction with LLMs leverages the LLM's capability to discern user intent and generate or modify outputs accordingly, researchers have proposed augmenting LLMs with enhanced reasoning abilities to expand their problem-solving capacity beyond basic inquiries such as mathematical problem-solving~\cite{lemmer_human-centered_2023} and argumentative writing~\cite{mirowski_co-writing_2023}.
The foundational approach to reasoning in text involves employing the Chain-of-Thought method~\cite{wei2023chainofthought}. This approach breaks down complex tasks into smaller, more manageable steps, using the output from previous steps to inform subsequent ones. While important and distinct from standard text-based conversational prompting, we found that this mode is primarily linked to UI design in HCI. We will provide more examples in Mode 2.5 in Section \ref{sec:human+UI+reasoning}.}

\begin{figure*}[!htbp]
    \centering
    \includegraphics[width=0.6\linewidth]{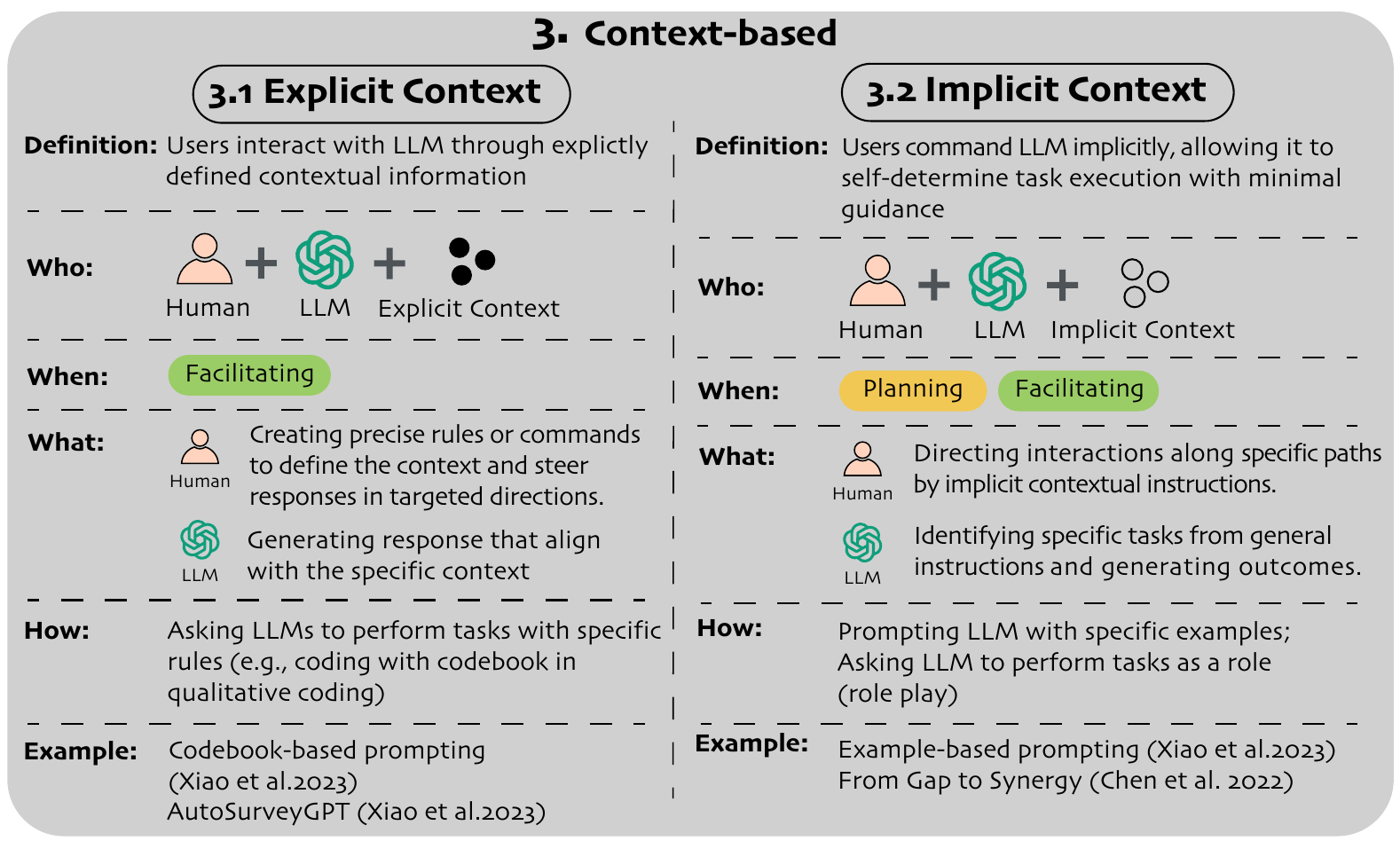}
    \caption{\context.}
    \label{fig:taxonomy3}
\end{figure*}

\subsubsection{\textbf{Mode 2: User Interface (UI)}}

Uer Interface (Figure \ref{fig:taxonomy2}) is a pivotal and practical means to enhance LLMs with advanced capabilities.

\paragraph{\textbf{Mode 2.1. UI for Structured Prompts Input}}
This approach enhances LLM inputs through a structured UI. For instance, distinct UI elements can be employed to input various components of a prompt, such as zero-shot, few-shot examples, and specific constraints. This approach ensures that each prompt is created easily and consistently, allowing users to concentrate on the key contents rather than spending time crafting a comprehensible prompt.  For instance, Jiang et al.\cite{jiang_promptmaker_2022} introduced PromptMaker, a tool that combines Prefixes, Settings, and Examples to create structured prompt inputs. Similarly, Dang et al.\cite{dang_choice_2023} developed UI variants that integrate user instructions with the standard prompting, enhancing more nuanced user-model interaction.

\paragraph{\textbf{Mode 2.2. UI for Varying Output}}
\label{sec:UI-varying-output}
UI design can further enhance LLM outputs by providing users with options to specify output formats and controls. These include selecting the size, picking the color, or choosing button layouts via the output interface. The aim is to enable LLMs to produce results that are not only more functional but also more complex, structured, and layered, providing greater depth and utility in the generated content. A typical example is GenLine, developed by Jiang et al.\cite{jiang_discovering_2022}, which enables users to generate CSS styles, such as button-style HTML code, and offers an interface for users to choose whether to accept the generated style. A variation of this tool is GenForm\cite{jiang_genline_2021}, which facilitates the structured generation of mixed outputs, including HTML, JavaScript, and CSS code, through a form interface. Contrasting with PromptMaker's focus on structuring prompt inputs through UI design, GenLine and GenForm prioritize transforming code generation into structured outputs for enhanced user consumption.

\begin{figure*}[!htbp]
    \centering
    \includegraphics[width=0.6\linewidth]{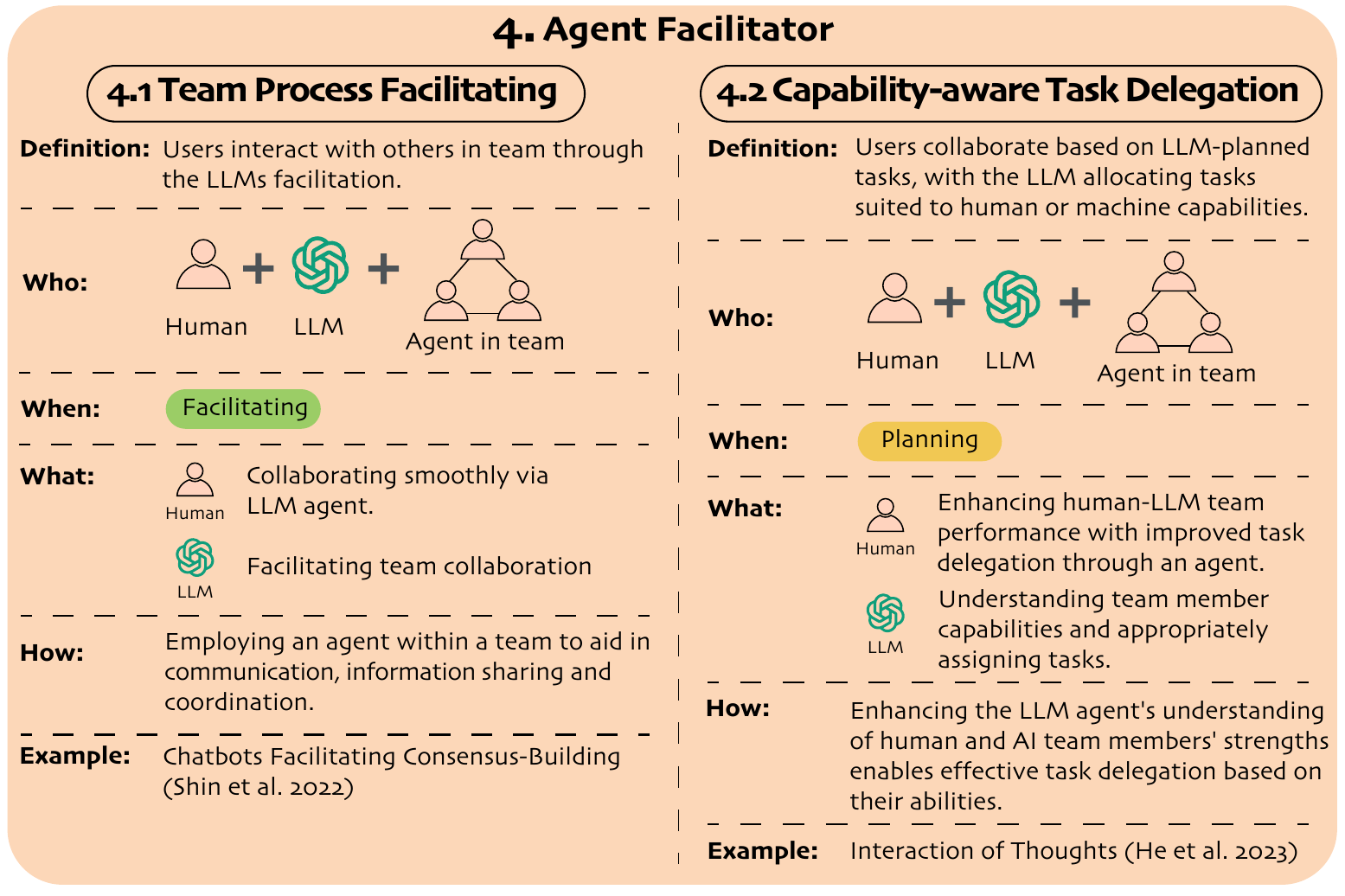}
    \caption{\agent.}
    \label{fig:taxonomy4}
\end{figure*}

\paragraph{\textbf{Mode 2.3. UI for Iteration of Interaction}}
UI design can significantly improve the iterative aspects of an interaction flow, incorporating features like debugging, error labeling, regenerating, and self-repairing. Such UI enhancements allow users to refine their original interaction flows, leading to improved final or intermediate outputs~\cite{liu_what_2023, brade_promptify_2023}. For instance, BotDesigner~\cite{zamfirescu-pereira_why_2023} aids users in refining human-LLM interactions, such as recipe conversations. It allows users to identify and label errors within the conversation via its interface and offers a "retry" button to regenerate the intermediate output, ensuring the integrity of the original interaction.

\paragraph{\textbf{Mode 2.4. UI for Testing of Interaction}}
UI design is employed to facilitate the testing of various prompt variations within an interaction flow. This capability is particularly useful for quickly prototyping complex artifacts, such as long writings, allowing users to experiment with and refine their creations with different inputs, prompts, and models. A typical example is VISAR~\cite{zhang2023visar}, which employs visual programming to give users control over the framework of argumentative writing and facilitates rapid prototyping of prompt ideas, enabling quick testing of writing organization. Similarly, Kim et al.~\cite{kim_cells_2023} introduced a new interface that allows end users to experiment with model configurations and inputs using object-oriented interaction.

\paragraph{\textbf{Mode 2.5. UI for Reasoning}} 
\label{sec:human+UI+reasoning}
Expanding upon the basic forms of reasoning augmentation in Mode 1.2 in Section \ref{sec:human+llm+reasoning}, a significant advancement involves incorporating direct manipulation through UI design into the Chain-of-Thought process. This approach allows users to actively participate in the reasoning sequence, providing immediate control at intermediate steps to alter the direction or nature of the reasoning. Users might seek not only to create the reasoning process but also to reorganize reasoning blocks in a manner that aligns with their unique thought processes. The objective is to tackle complex tasks, yet with enhanced control, customization, and precision. This is facilitated by employing visual programming techniques, such as chain designs~\cite{wu_promptchainer_2022, wu_ai_2022, arawjo_chainforge_2023} and mind maps~\cite{jiang2023graphologue, suh2023sensecape, zhang2023visar}, enabling a more interactive and user-defined reasoning framework. 
\added{While there may be some overlap with other modes, such as \textit{Mode 2.2 UI for Varying Output}, we classify this as a distinct submode due to its unique blend of reasoning and UI design. }

\subsubsection{\textbf{Mode 3: Context-based}}
Context-based mode focuses on augmenting the system with specific contextual understandings (Figure \ref{fig:taxonomy3}). Although this mode is less widespread and dominant compared to \standard and \uiinterface, it may serve as a key direction in design, potentially inspiring further work within two distinct approaches—explicitly defined rules/contexts or implicit contexts.

\paragraph{\textbf{Mode 3.1. Explicit Context}}
Augmenting LLMs with an explicit context involves prompting to process and respond to information based on predefined dimensions or contextual rules.  For instance, codebook-centered prompting~\cite{xiao_supporting_2023} enables researchers to provide LLMs with a codebook for qualitative analysis, outlining specific design patterns in data. 
Similarly, AutoSurveyGPT~\cite{xiao_autosurveygpt_2023} enables LLMs to automatically scan abstracts and extract keywords based on pre-defined commands and rules.

\paragraph{\textbf{Mode 3.2. Implicit Context}}
In contrast, augmenting LLMs with implicit context involves providing them with limited or general queries and commands, and then expecting them to infer how to perform tasks based on a few examples or by detecting the underlying context~\cite{srinivasan_snowy_2021}. This approach demands the LLM's interpretation of user intent and subtle signals, fostering a nuanced grasp of context for tasks needing profound contextual insight. Typical examples include \textbf{role play} and \textbf{example-based prompting}. In role play, users enhance LLMs' output quality by assigning them specific roles, such as a chatbot capable of reflective thinking~\cite{kumar_exploring_2023}. This method involves inputting detailed information profiles, such as characteristics and areas of expertise, to closely simulate real expert knowledge. Further enhancements are possible through structured UI design, enabling the establishment of precise characteristics under user control, thereby boosting the LLM's role-play efficacy. Similarly, example-based prompting entails providing LLMs with a handful of relevant examples that clearly demonstrate the expected input and output~\cite{xiao_supporting_2023}, with the LLM tasked to independently discern the underlying rationale.

\subsubsection{\textbf{Mode 4: Agent Facilitator}}
The agent facilitator mode (Figure \ref{fig:taxonomy4}) focuses on enhancing team dynamics and performance through LLMs acting as facilitators.

\paragraph{\textbf{Mode 4.1. Team Process Facilitating}}
In this approach, LLMs are used to streamline and facilitate the team's interaction process, particularly during the \textit{facilitating} phase. 
This is typically achieved by integrating an agent within the team that aids in communication~\cite{duan_bridging_2021}, decision-making~\cite{zheng_competent_2023}, information sharing, and coordination. By smoothing out the interaction process, the LLM ensures that the team's workflow continues seamlessly and effectively, enhancing collaboration and productivity. 

\paragraph{\textbf{Mode 4.2. Capability-aware Task Delegation}}
This approach involves augmenting LLMs with the ability to recognize the unique capabilities of different team members. The primary goal is to leverage the diverse skills of team members to optimize overall team performance, ensuring that tasks are assigned in a way that maximizes each member's contribution and efficiency. For instance, in response to questions like `Can AI perform well?'~\cite{shi_retrolens_2023}, the LLM can decide whether to assign tasks to each member of a group. This is typically crucial during the \textit{planning} phase and can also be relevant in the \textit{facilitating} phase.

%% file: paper/5_limitation.tex
\section{Discussion}

\subsection{Two Potential Applications of Interaction Modes Taxonomy}

\added{In this paper, we describe a taxonomy specifically focused on possible interaction modes between human and LLMs. The goal is to empower users to tackle complex tasks by utilizing LLMs beyond the default conversational prompting paradigm. Similar to other software taxonomies, such as software design patterns~\cite{gamma1995design, schmidt2013pattern} and catalogs for prompt engineering~\cite{white2023prompt}, our taxonomy aims to assist software designers in at least two important ways. } 

\added{First, by offering a high-level, systematic, and multi-dimensional understanding of interaction modes, \textbf{this taxonomy empowers users to swiftly understand when and how to implement a specific mode, identify the stakeholders involved, and consider relevant factors}, thereby enhancing their system design and facilitating the evaluation of potential improvements. For instance, in brainstorming sessions, users can utilize specific interaction modes from the taxonomy to propose and refine details of their system design ideas. Furthermore, after proposing the primary system, this taxonomy can serve as a checklist for their designs, prompting critical questions such as, "Have I overlooked any crucial steps or details?" and "Does my design have any precedents?"}

\added{Second, \textbf{the taxonomy can unveil new possibilities previously unconsidered}. One approach is through the adoption of novel interaction modes that, while initially overlooked, prove to be invaluable. For instance, iterating on an established interaction by identifying errors and retrying (\textit{Mode 2.4: UI for Testing of Iteration}), or utilizing an LLM-based agent to enhance team interactions (\agent).
Additionally, users can discover new opportunities by merging elements from various modes, such as UI design, reasoning, context, and so on. Several examples include ``\textit{human+LLM+role play+UI for input/output}" during facilitation; ``\textit{human+LLM+few-shot examples+UI for input}" during facilitation; and ``\textit{human+LLM+explicit constraints+UI for input/output}" during planning and facilitation. These combinations foster innovative approaches but have not yet been explored in the existing research in our reviewing.}

\added{In addition to the ways of applications, we believe our taxonomy can be used in many tasks. While the core of our taxonomy--prompts--as a form of natural language, initially find their primary application in writing and coding tasks, we observe their application broadening to encompass more tasks that integrate prompts into the system's core. For instance, in the realm of image generation, although the system's objective is to produce images, it still necessitates natural language prompts as the initiation point. Hence, it is plausible to anticipate that the taxonomy of interaction modes could serve as a useful tool for designing prompts across various domains, such as image and video generation. Overall, our vision with this taxonomy is to think of the augmentation of prompts of LLMs as a new form of "software" for users to interact with hardware and encapsulate data for efficient task execution.}

\subsection{Limitations and Future Work}

\added{As highlighted in the title, the taxonomy presented herein represents an initial endeavor, which we intend to refine continuously. We foresee its expansion to encompass additional LLM interaction modes likely to emerge in the near future. Moreover, it is important to note that many current classifications in our taxonomy are not absolute, given the slight overlap between some categories and the potential for misapplication.
}

\added{Looking forward, we believe that it will be especially valuable to extend the taxonomy to explicitly include different kinds of tasks and different design spaces. For instance, we believe that the capabilities and interaction modes needed to \textit{creating} tasks will likely be systematically different from the capabilities and modes needed to \textit{deciding} tasks. To do that, we plan to extend our literature review to additional venues from diverse fields such as ACL, EMNLP, NAACL, TACL, and broader HCI venues like TOCHI, C\&C, DIS, and even arXiv.}

%% file: paper/6_conclusion.tex
\section{Conclusion}

In this paper, we adopt an HCI perspective to explore  examine interaction modes—patterns we can leverage to enhance LLMs' capabilities through diverse human-LLM interaction designs. Our literature review within major HCI venues has led us to identify distinct phases of human-LLM interaction flow, and iteratively developed a taxonomy that encapsulates four key interaction modes in human-LLM interaction. This taxonomy provides a valuable tool for systematically understanding and analyzing the evolving landscape of human-LLM interaction and collaboration. It guides the design of human engagement with LLMs in increasingly complex and nuanced ways.

\begin{acks}
This research is supported by the National Research Foundation (NRF), Prime Minister’s Office, Singapore under its Campus for Research Excellence and Technological Enterprise (CREATE) programme. The Mens, Manus, and Machina (M3S)
is an interdisciplinary research group (IRG) of the Singapore-MIT Alliance for Research and Technology (SMART) centre.
\end{acks}

%% file: paper/appendix.tex
\appendix

\begin{table*}[!htbp]
\caption{The table shows all the papers included in the taxonomy.}
\begin{tabular}{lll}
\hline
Taxonomy & Subcatagory & Citations \\ \hline
\multirow{2}{*}{Mode 1. Standard Prompting} & Mode 1.1. Text-based Conversational Prompting & \begin{tabular}[c]{@{}l@{}}\cite{zhu_leveraging_2023} \cite{jung_toward_2023} \cite{ross_programmers_2023} \cite{jo_understanding_2023}  \\\cite{lee_coauthor_2022} \cite{zhang_storybuddy_2022} \cite{lc_designing_2021} \cite{xiao_let_2021} \\ \cite{bursztyn_developing_2021} \cite{reynolds_prompt_2021} \cite{buschek_impact_2021}  \cite{mukherjee_impactbot_2023}\end{tabular} \\ \cline{2-3} 
 & Mode 1.2. Text-based Conversational Prompting with Reasoning & \cite{wu2022ai} \cite{lemmer_human-centered_2023} \\ \hline
\multirow{5}{*}{Mode 2. User Interface (UI)} & Mode 2.1. UI for Structured Input & \begin{tabular}[c]{@{}l@{}}\cite{kim_facilitating_2022} \cite{liu_opal_2022} \cite{angert_spellburst_2023}\cite{louie_expressive_2022} \\ \cite{dang_choice_2023} \cite{wang_popblends_2023} \cite{dang_how_2022} \cite{jiang_promptmaker_2022} \\ \cite{osone_buncho_2021}\end{tabular} \\ \cline{2-3} 
 & Mode 2.2. UI for Varying Output  & \begin{tabular}[c]{@{}l@{}} \cite{jiang_genline_2021} \cite{ito_use_2023} \cite{bhat_interacting_2023} \cite{yuan_wordcraft_2022} \\ \cite{narechania_diy_2021} \cite{zhou_interactive_2021} \cite{mcnutt_design_2023} \cite{gebreegziabher_patat_2023} \\ \cite{petridis_promptinfuser_2023} \cite{jiang_discovering_2022} \cite{kim_stylette_2022} \cite{chung_talebrush_2022} \\ \cite{brachman_follow_2023} \cite{mirowski_co-writing_2023} \end{tabular}  \\ \cline{2-3} 
 & Mode 2.3. UI for Iteration of Interaction & \begin{tabular}[c]{@{}l@{}}\cite{cuadra_my_2021}  \cite{zhao_narratron_2023} \cite{brade_promptify_2023}  \cite{dang_beyond_2022} \\ \cite{jung_interactive_2023} \cite{gu_augmenting_2023}  \cite{liu_what_2023}  \cite{zamfirescu-pereira_why_2023}\end{tabular} \\ \cline{2-3} & Mode 2.4. UI for Testing of Interaction & \cite{kim_cells_2023} \cite{zhang2023visar} \cite{evirgen_ganzilla_2022} \\ \cline{2-3} 
 & Mode 2.5. UI for Reasoning & \begin{tabular}[c]{@{}l@{}}\cite{zhang_towards_2023}     \cite{wu_promptchainer_2022} \cite{jiang_graphologue_2023}  \cite{suh_sensecape_2023} \\ \cite{arawjo_chainforge_2023} \cite{wang2023enabling} \end{tabular} \\ \hline
\multirow{2}{*}{Mode 3. Context-based} & Mode 3.1. Explicit Context & \begin{tabular}[c]{@{}l@{}}\cite{xiao_autosurveygpt_2023} \cite{pu_semanticon_2022} \cite{wu_scattershot_2023}  \cite{xiao_supporting_2023}\end{tabular} \\ \cline{2-3} 
 & Mode 3.2. Implicit Context & \begin{tabular}[c]{@{}l@{}}\cite{chen_gap_2023} \cite{srinivasan_snowy_2021} \cite{kumar_exploring_2023} \cite{srinivasa_ragavan_gridbook_2022}  \\ \cite{xiao_supporting_2023} \cite{fan_human-ai_2022} \end{tabular} \\ \hline
Mode 4. Agent Facilitator & Mode 4.1. Team process facilitating & \cite{duan_bridging_2021} \cite{shin_chatbots_2022} \cite{zheng_competent_2023} \cite{purohit_chatgpt_2023}\\ \cline{2-3} 
 &  Mode 4.2. Capability-aware Task Delegation & \begin{tabular}[c]{@{}l@{}} \cite{lee_exploring_2021}  \cite{he_interaction_2023}  \cite{shi_retrolens_2023} \cite{lai_human-ai_2022}\end{tabular} \\ \hline
\end{tabular}
\label{tab:paper_classification}
\end{table*}

\begin{table*}[!htbp]
\caption{The primary version of taxonomy in Section \ref{sec:primary_taxonomy}.}
\label{tab:primary_taxonomy}
\scalebox{0.6}{\begin{tabular}{|l|l|l|l|l|l|}
\hline
\textbf{Interaction Modes} & \textbf{Who} & \textbf{When} & \textbf{Wow} & \textbf{Definition} & \textbf{Example} \\ \hline
\multirow{2}{*}{UI+prompts} & \multirow{2}{*}{\begin{tabular}[c]{@{}l@{}}User,\\ LLMs+UI\end{tabular}} & Proposing & \begin{tabular}[c]{@{}l@{}}Consistent and comprehensive \\ input prompts\end{tabular} & \begin{tabular}[c]{@{}l@{}}Using an interface design to structure \\ the input of zero-shot, few-shot. Propose \\ interface that supports the prompts to \\ be step-by-step evolved/organize \\ prompts/input prompts structurely\end{tabular} & \begin{tabular}[c]{@{}l@{}}Design customized UI to get customized prompts: \\ (1) Complete the sentence. \\ (2) Complete the sentence and \textless{}user\_instruction\textgreater{}.\\      \\ Structured interface that allows users to input \\ few-shot examples consistently\end{tabular} \\ \cline{3-6} 
 &  & Iteration & \begin{tabular}[c]{@{}l@{}}More   rounds of prompting, \\ for reflection and improvement \\ of the quality\end{tabular} & \begin{tabular}[c]{@{}l@{}}Iteration of the prompting /conversational \\ process: (1) allows users to create an \\ LLM-based chatbot solely through prompts, \\ and (2) encourages iterative design and \\ evaluation of effective prompt strategies.\end{tabular} & \begin{tabular}[c]{@{}l@{}}1. Defning a “baseline” chatbot prompt template\\ 2. Assessing what the baseline bot is capable of\\ 3. Identifying errors.\\ 4. Debugging\\ 5. Evaluating the new prompt locally.\\ 6. Evaluating the new prompt globally.\\ 7. Iteration.\end{tabular} \\ \hline
\multirow{2}{*}{\begin{tabular}[c]{@{}l@{}}Conversational and \\ Step-by-step\end{tabular}} & \multirow{2}{*}{\begin{tabular}[c]{@{}l@{}}User,\\ LLMs+reasoning\end{tabular}} & \begin{tabular}[c]{@{}l@{}}Proposing   \\ Iteration\end{tabular} & \begin{tabular}[c]{@{}l@{}}Improve prompt quality, \\ iterate the results\end{tabular} & \begin{tabular}[c]{@{}l@{}}Human gives input and get several intermediate \\ results from LLMs and then do editing/give \\ feedback to iterate the results\end{tabular} & \begin{tabular}[c]{@{}l@{}}Low-codeLLM: VisualProgramming overLLMs:\\ “What It Wants Me To Say”: Bridging the \\ Abstraction Gap Between End-User Programmers \\ and Code-Generating Large Language Models\end{tabular} \\ \cline{3-6} 
 &  & Proposing & \begin{tabular}[c]{@{}l@{}}Acomplish reasoning on \\ complex tasks.\end{tabular} & \begin{tabular}[c]{@{}l@{}}For   complex reasoning tasks, split the \\ tasks into subtasks.\end{tabular} & \begin{tabular}[c]{@{}l@{}}Chain of thought, Self-Consistency, \\ Automatic Reasoning and Tool-use (ART).\\ Decomposed Prompting : A MODULAR \\ APPROACH FOR SOLVING COMPLEX TASKS\end{tabular} \\ \hline
Role Play & \begin{tabular}[c]{@{}l@{}}User,\\ LLMs+role/persona\end{tabular} & Proposing & \begin{tabular}[c]{@{}l@{}}Get responses like some \\ pre-defined persona, \\ improve the quality of results\end{tabular} & Setting the characteristics/role/persona & e.g., Ask GPT to act as a proofreader \\ \hline
Context based & \begin{tabular}[c]{@{}l@{}}User,\\ LLMs + context\end{tabular} & Proposing & \begin{tabular}[c]{@{}l@{}}Drive   responses into \\ specific directions\end{tabular} & \begin{tabular}[c]{@{}l@{}}Ask GPT to generate outputs from \\ different dimensions\end{tabular} & \begin{tabular}[c]{@{}l@{}}deductive/inductive, \\ example-centered/codebook-centered\end{tabular} \\ \hline
\end{tabular}}
\end{table*}